# Phase changes in delay propagation networks

Seddik Belkoura, Massimiliano Zanin
The INNAXIS Foundation & Research Institute
Madrid, Spain
{sb, mz}@innaxis.org

*Abstract*— The analysis of the dynamics of delays propagation is one of the major topics inside Air Transport Management research. Delays are generated by the elements of the system, but their propagation is a global process fostered by relationships inside the network. If the topology of such propagation process has been extensively studied in the literature, little attention has been devoted to the fact that such topology may have a dynamical nature. Here we differentiate between two phases of the system by applying two causality metrics, respectively describing the standard phase (*i.e.* propagation of normal delays) and a disrupted one (corresponding to abnormal and unexpected delays). We identify the critical point triggering the change of the topology of the system, in terms of delays magnitude, using a historical data set of flights crossing Europe in 2011. We anticipate that the proposed results will open new doors towards the understanding of the delay propagation dynamics and the mitigation of extreme events.

## I. Introduction

In the last decade, the study of complex systems has shifted from the reductionist hypothesis towards an information processing approach, *i.e.* how information is distributed among, combined with, and modified by the different elements of the system [1]. Not only the constituting elements are important, but also their relationships; and understanding the complete system becomes tantamount to analyzing how information is processed within, and exchanged between, the individual elements. Examples span from the study of financial markets [2] to the human brain [3].

This information processing approach has recently been adapted to air transport, and specifically to the study of delays propagation. While delays are generated at a local scale (*i.e.* by individual aircraft), their propagation is a process that lies on a more global scale. Local metrics, as for instance "delay multipliers", are useful to understand how delays are generated [4, 5]. Nevertheless, shedding light on how delays propagate, as the result of interactions between flights and airports, requires a more systemic approach, as the one provided by complex networks [6-8].

In recent years, complex networks have been used to characterize the structures created by delays propagation – see, for instance, [9-12]. Of special interest is the use of *functional networks*, *i.e.* networks in which nodes (in this case, airports) are connected by a link if a delay propagation process has been identified within the corresponding operational data. In other words, suppose one is to identify if a delay propagation happened between airports *A* and *B*. The solution entails obtaining two time series, for instance the average hourly delays in both *A* and *B*; for then calculating the presence of a correlation (or of a causality) between both time series. If a causality $A \to B$ is detected, one can then conclude that the delays observed in *B* are (partly) the result of the delays observed in *A*.

The use of functional networks for the understanding of delays propagation presents some important advantages. First, networks yield a global view of the propagation process, beyond the dynamics of individual flights and airports. Second, they are reconstructed using only real operational data, and not *a priori* information; on the contrary, models of delays propagation can be useful, but their results are as good as the models themselves – if some processes are not correctly modeled, results may be misleading. Third, they are able to identify indirect propagations, *i.e.* propagations of delays between pairs of airports not connected by a direct fly.

On the other hand, functional networks also entail an important problem: all available data are used in their reconstruction. In other words, if a causality is detected between two airports, it means that "on average" a propagation process is present. Let us state this the other way around: if an abnormal delay propagation only appears under certain conditions, such relation may be smoothed by the average process, thus yielding no statistically significant causality.

In this contribution, we tackle this problem by comparing the results obtained through two causality metrics. First, the well-known Granger Causality (GC) [13-14], a metric that assesses whether a time series can be used to forecast a second one. This analysis is performed over all the available data, to provide a quantification of the average causality relation between the series. As such, GC suffers from the aforementioned problem of statistical smoothing. Second, a recently proposed causality of extreme events [15], which detects causal relationships between time series by only

This work is co-financed by EUROCONTROL acting on behalf of the SESAR Joint Undertaking (the SJU) and the EUROPEAN UNION as part of Work Package E in the SESAR Programme. Opinions expressed in this work reflect the authors' views only and EUROCONTROL and/or the SJU shall not be considered liable for them or for any use that may be made of the information contained herein.

considering abnormal values, and the statistics of how such values are propagated – thus disregarding the underlying normal dynamics. These two metrics allow us to respectively define two different phases of the system: (*i*) the standard delays propagation process, *i.e.* the one taking place during normal days; and (*ii*) a disrupted one, observed when some events trigger abnormally high delays in the system. By using complex networks theory, we compare both phases, and show how some airports undergo important changes in their role. Furthermore, we estimate the threshold for the phase change to occur, *i.e.* the amount of additional delay that is needed to change the behavior of the system.

Beyond this introduction, this contribution is organized as follows. Section II describes the mathematics of the two causality measures here considered, and discusses the differences in their approaches. Section III introduces the data set used in the analysis, and the pre-processing executed to extract the delay time series. Afterwards, Section IV presents the obtained results, both in terms of delay propagation networks and of the additional delay needed to trigger a phase change. Finally, Section V draws some conclusions and outlines future steps.

## II. DETECTING CAUSALITY IN DELAY DATA

### A. Granger Causality

The Granger Causality test [13-14] is an extremely powerful tool for assessing information exchange between different elements of a system, and understanding whether the dynamics of one of them is led by the other(s). In the case at hand, it will help understand if the delays observed at one airport are caused by the dynamics of other airport(s).

Although Granger Causality has largely been applied to economic problems [16], as it was originally developed by the economy Nobel Prize winner Clive Granger [13], it has recently received significant attention in the analysis of biomedical [17-19] and air transport [20, 21, 22] data.

The GC test is based on two very simple ideas, which take the form of two axioms:

- Causes must precede their effects in time.
- Information relating to a cause's past must improve the prediction of the effect above and beyond information contained in the collective past of all other measured variables (including the effect).

Therefore, an airport $A$ is considered to "Granger-cause" another airport $B$ if the inclusion of past values of the delays time series of $A$ can improve the process of forecasting the values of the delay time series in $B$. In other words, the future evolution of delays in airport $B$ also depends on the past values of delays in $A$.

In mathematical terms, suppose that the delay dynamics of two airports $A$ and $B$ can be expressed as stationary time series[1]. Then $A$ "Granger-causes" $B$ if:

$$\sigma^2(B|U^-) < \sigma^2(B|U^- \setminus A^-). \qquad (1)$$

$\sigma^2(B|U^-)$ denotes the variance of the residuals of predicting the time series $B$ using the information of the entire universe $U$ available at present time, and $\sigma^2(B|U^- \setminus A^-)$ the corresponding variance if $A$ is excluded from this universe. The statistical significance of the inequality is then expressed as a *p*-value, calculated through a standard F-test.

One of the major advantages of the GC test is that it is able to discriminate causalities from simple correlations. In the latter case, as the time series would have similar values, one of them cannot convey useful information for the forecast of the other. Yet, claims of causality from (multiple) bivariate time series should always be taken with caution, as true causality can only be assessed if the time series contain all possible relevant information and sources of activities for the problem [23], a condition that real-world experiments can only rarely comply with [24].

### B. Causality of extreme events

If Granger Causality has extensively been used in the analysis of real-world data, it has also been recognized that it presents several drawbacks [25]. From the point of view of a delay propagation analysis, two have to be highlighted. First, this metric is linear, in the sense that it assesses the presence of linear couplings between the time series – while it is well known that delays usually propagate in a non-linear fashion. Second, it looks for causalities over the whole time series; if a delay propagation only occurs in a specific time window, for instance because of some specific traffic conditions, such propagation would be smoothed and averaged, yielding a low statistical significance.

In order to overcome these limitations, we here consider a recently proposed metric for the detection of causality in extreme events [15]. It is based on identifying extreme events in two time series, *i.e.*, in the problem at hand, situations in which higher than expected delays are recorded for two airports; and on studying the statistics of how they co-occur.

Suppose once again two airports, $A$ and $B$, for which one ought to detect if a causality $A \to B$ is present. If such causality is indeed present, most (ideally, all) of the extreme events of $A$ should correspond to extreme events of $B$; in other words, abnormal delays in $A$ would propagate to $B$ in a finite time. At the same time, one would expect that extreme events of $B$ only partially correspond to events of $A$; as $B$ does not causes $A$,

---

[1] "Stationary", in time series analysis, refers to the fact that a time series has no explicit relation with time, *e.g.* no seasonality or trends can be detected. Refer to Section III for a discussion on how delay time series can be made stationary.

delays may appear in the former without being propagated to the latter.

Let us denote by $p_1$ the probability that an extreme event in $B$ also corresponds to an extreme event in $A$; and by $p_2$ the probability that an extreme event in $A$ corresponds to an extreme event in $B$. The previous discussion suggests that a causality $A \rightarrow B$ will be present if $p_1 > p_2$. The statistical significance of such relation can be quantified through a binomial two-proportion z-test:

$$z = \frac{p_1 - p_2}{\sqrt{\hat{p}(1-\hat{p})\left(\frac{1}{n_1} + \frac{1}{n_2}\right)}},$$

being $n_1$ and $n_2$ the number of events associated to $p_1$ and $p_2$, and $\hat{p} = (n_1 p_1 + n_2 p_2)/(n_1 + n_2)$. The corresponding $p$-value can be obtained through a Gaussian cumulative distribution function.

One last issue shall be discussed, *i.e.* how the extreme events are defined. On one hand, there is no *a priori* way of defining them that does not introduce some bias; that is, choosing an arbitrary threshold, *e.g.* the top 1% of all delay values, is not readily justifiable with the available data. On the other hand, when comparing the extreme events of two airports, it seems reasonable to expect them to have different magnitudes: 10 minutes of delay may be unexpected in one airport, while perfectly normal in another one. Taking into account these two issues, the algorithm for the causality for extreme events resorts to a full search of the parameter space; it tries every combination of thresholds for each pair of airports, and selects the one yielding the best statistical significance.

More details about the properties of this metric, and about the software implementation, can be found in [15].

*C. A comparison of the two causalities*

Let us recapitulate the main differences between the two causality metrics here considered, as these will be the foundation for understanding the results presented in Section IV.

On one hand, the Granger Causality states that an airport $A$ "Granger-causes" another airport $B$ if the past of $A$ contains information that helps predicting the future of $B$, above the information already contained in $B$. In general terms, when this condition is satisfied, one can say that there is an "information flow" from $A$ to $B$.

On the other hand, the newly proposed causality metric of extreme events focuses solely on the co-movements between the right tails of the delay distributions, without taking into account the remaining data. Thus, airport $A$ causes airport $B$ if extremely high delays in $A$ trigger abnormal delays in $B$.

It is thus simple to describe the main difference between these two metrics. The Granger Causality detects when a delay in airport $A$ is systematically "transferred" to the airport $B$ under normal conditions, while the extreme events causality analyses the system only when it is in a disrupted state. This allows to describe two different phases of the delay propagation process: the normal flow of information between the airports, as yielded by the Granger causality; and the disrupted state, triggered by unexpected high delays, described by the causality of extreme events.

III. DATA SET DESCRIPTION

The time series describing the delay dynamics of European airports have been extracted from the Flight Trajectory (ALL-FT+) data set, as provided by the EUROCONTROL PRISME group. This data set includes information about planned and executed trajectories for all flights crossing the European airspace, with an average resolution of 2 minutes. The data set covers the period from 1st March to the 9th of August 2011. A subset has been extracted, composed of the 100 busiest airports in 2011, including a total of 2.8 million flights.

From this data set, a time series for each airport has been extracted, corresponding to the average landing delay within one hour time windows. The delay of each flight is calculated as the difference between the real and the planned landing time – note that this can yield positive or negative average delays.

The result is a matrix $D$ of size 100 x 3888, whose element $d_{ij}$ corresponds to the average delay of airport $i$ (with $i \in \{1 ... 100\}$) at time $j$ ($j \in \{1 ... 3888\}$, the number of hours covered by the data set). These time series present important trends – as, for instance, it is known that delays tend to appear during peak hours, and almost disappear at night. In order to fulfill the stationarity requirement of the Granger Causality, the data have been pre-processed to eliminate trends as follows:

$$\forall i,j, \qquad d_{ij} = \frac{d_{ij} - \overline{d_{il}}}{\sigma^2(d_{il})}, \qquad l = j \pm 24k, k \in \mathbb{N}$$

$\overline{d_{il}}$ and $\sigma^2(d_{il})$ respectively corresponding to the average delay and the standard deviation encountered in the airport $i$, and $l$ running over all data with the same hour. In other words, $d_{ij}$ now corresponds of the Z-Score of the delays encountered at the same hour at the same airport, during different days. High positive (and negative) values of $d_{ij}$ indicate the presence of a delay larger (or smaller) than expected, and are thus independent from the original magnitude of the delay.

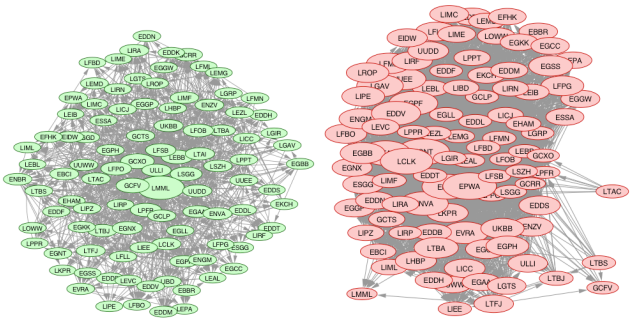

Figure 1. Networks obtained for the Granger (Left) and extreme events (Right) causalities. Node sizes are proportional to the number of causality connections generated at each airport.

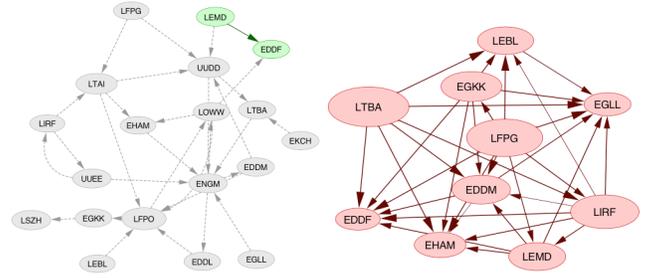

Figure 2. Networks of the top 10 airports in number of operations, for the Granger (Left) and extreme events (Right) causalities. In the former case, only two airports are part of the top 10 network (LEMD and EDDF, in green); for the sake of clarity, airports of the top 20 are also drawn, in light grey.

IV. DELAY CAUSALITY AND PHASE CHANGES

*A. Delay causality networks*

When the two causality metrics introduced in Section II are applied to the described European data, *i.e.* by assessing the presence of a causality between each pair of airports, the results are the two networks represented in Fig. 1. The links represented are those that are statistically significant (*p*-value or *F*-value < 0.01), and for which the presence of a confounding effect can be discarded[2]. For the sake of clarity, all results related to the Granger Causality network are represented in green, those to the causality of extreme events in red.

A simple visual inspection suggests that both networks have a different structure, and specifically that the extreme events one has a much higher link density (total number of connections). This same fact is highlighted in Fig. 2, where only the sub-network of the 10 most important airports is drawn. In order to quantify such difference, Tab. I reports the values of some relevant topological metrics for both networks (more details on these metrics can be found in the literature, *e.g.* [6-8]):

- *Link density*, defined as the number of links present in the network, divided by the number of all possible links.

- *Transitivity*. Measures of the degree to which nodes in a graph tend to cluster together and to form triangles. It is calculated as the number of closed triplets (or triangles) over the total number of triplets (groups of three nodes connected by links).

- *Efficiency*. Measures how efficiently the network is able to transmit information (in this case, delays) between its nodes. It is defined as the inverse of the harmonic mean of the distances between pairs of nodes.

- *Assortativity*. Positive values of this metric indicate that nodes tend to connect to other nodes with the same number of connections. On the other hand, negative values suggest that, on average, highly connected nodes are mostly linked to isolated ones.

- *Diameter*. The greatest distance between any pair of vertices.

- *Information Content* (IC) [26]. This metric assesses the presence of mesoscale structures, *i.e.* structures created by small groups of nodes, by evaluating the information lost when pairs of nodes are iteratively merged together. High values of Information Content indicate a random-like structure; conversely, small values suggest a non-trivial topology.

Some metrics are not directly comparable, as they depend on the link density of the network; in these cases, the corresponding Z-Score is reported in parenthesis, defined as the deviation with respect to the value expected in an ensemble of random equivalent (same number of nodes and links)

TABLE I. RESUME OF TOPOLOGICAL METRICS

| Metric | GC Network | E. E. Network |
|---|---|---|
| Link density | 0.1820 | 0.6479 |
| Transitivity | 0.2703 (-0.2227) | 0.7382 (-0.2421) |
| Efficiency | 0.5910 (1.7912) | 0.8238 (0.0423) |
| Assortativity | -0.2227 | -0.2421 |
| Diameter | 3 | 2 |
| Information Content | 0.9353 | 0.2683 |

---

[2] In the case of the extreme events causality, this entails discarding pairs of airports for which the $A \rightarrow B$ and $B \rightarrow A$ tests are both significant. See [15] for further details.

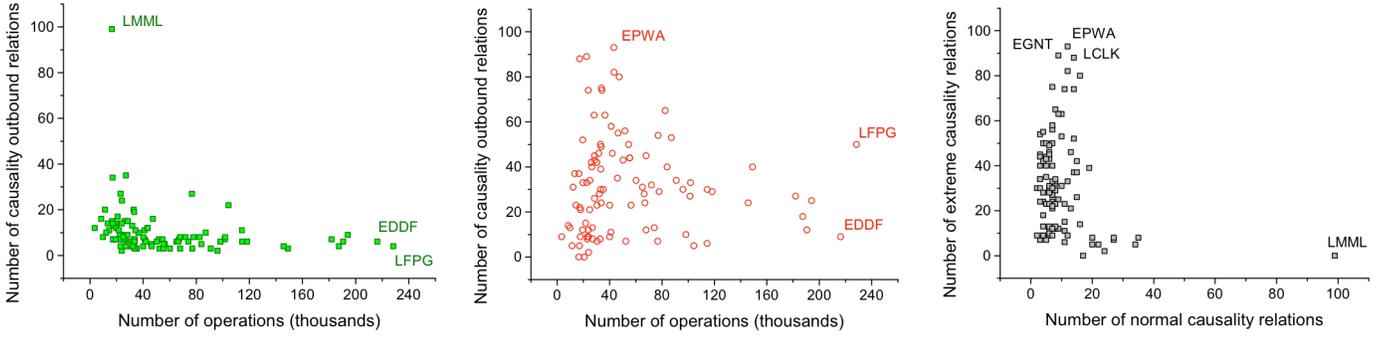

Figure 3. Relations between the number of causality links generated by each airport, and its volume of traffic. See main text for details.

networks:

$$Z - Score = \frac{m - \overline{r_m}}{\sigma_{r_m}},$$

$m$ being the value of the metric, $\overline{r_m}$ and $\sigma_{r_m}$ respectively the average value and standard deviation in the random ensemble.

It can be appreciated that the Granger Causality network has higher efficiency (in terms of Z-Score); thus, in spite of having fewer connections for propagating delays, such propagation is more effective. On the other hand, the network corresponding to extreme events is not that efficient, in spite of having a non-trivial topology (see the low value of the IC); this suggests that delays are mainly propagated locally, *i.e.* in small groups of nodes, depending on the type of events generating abnormal delays.

Fig. 3 presents some more results regarding the way links are organized around different airports. Specifically, Fig. 3 Left and Center depicts the number of connections (*i.e.* the number of causality relations generated) as a function of the number of operations recorded by each airport, respectively for the Granger and extreme events causalities. In both cases, there is an inverse relation, which is more evident in the latter case (the best fit being an exponential decay, R-Square of 0.066). Large airports are thus causing less delay propagation, probably because they can benefit from more buffers (*e.g.* stand-by aircraft, multiple runways to compensate for changes in wind direction, *etc.*); on the other hand, delays generated at small airports are readily propagated throughout the network. Similar findings have been reported in the literature, for the propagation of normal delays [5, 9, 27].

Fig. 3 Right depicts a comparison of the importance of airports in terms of propagation links, comparing the normal *vs.* disrupted phases. No clear relation can be observed: therefore, airports that may not propagate delays under normal conditions may suddenly become central when abnormal events appear. The opposite may also happen, as it is exemplified by the LMML airport (Malta International Airport), the most important airport in the normal phase, which looses most of its importance in the disrupted regime. The explanation of its behavior is two-fold. First, LMML is the smallest airport considered in this study, and thus the most subject to statistical fluctuations. Second, being a touristic destination, this airport is very well connected. Approximately 35 airlines ensure its link with a large number of European airports (including most of the airports in this study); in comparison, the LFOB airport (Beauvais-Tillé), 99[th] airport in term of passenger flow, is connected by only 4 airlines to less than 25 airports. This may imply that standard delays are

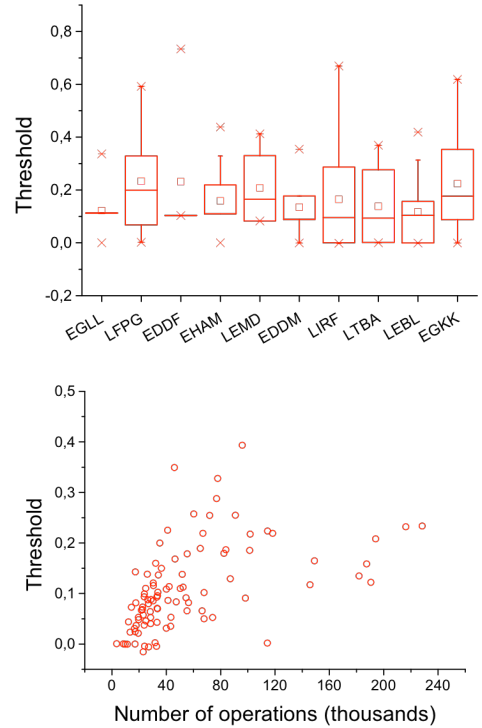

Figure 4. (Top) Box plot of the thresholds for the phase change, for the 10 airports with most operations. (Bottom) Average airport threshold as a function of the number of operations.

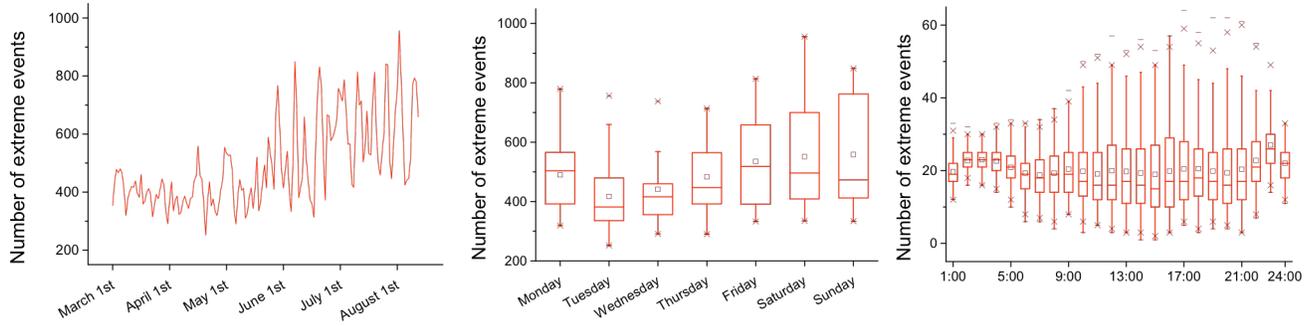

Figure 5. Evolution of the number of extreme events through time (Left), and as a function of the day of the week (Center) and of the hour of the day (Right).

propagated all over Europe, but that, at the same time, extreme events are most limited to the island.

*B. Phase transition*

As previously seen, the networks obtained by applying the two causality metrics are qualitatively and quantitatively different. They describe two phases of the network: the flow of delays under normal circumstances and the disrupted dynamics due to abnormal events. It is therefore of interest to look at when the transition between those two phases happens.

As explained in Section II-B, the causality of extreme events is based on the detection of shocks propagation between time series. This involves the tuning of a threshold, unique for each pair of airports, to define which event is to be considered as extreme. The statistical analysis of these thresholds can then be used to understand the magnitude of disturbances needed for a change in the phase.

Fig. 4 Top presents a box plot with the distribution of the thresholds for the 10 busiest airports. When considering the average (horizontal line in each box), one obtains the average additional delay (in terms of Z-Score) necessary for each airport to enter in the abnormal phase, *i.e.* the surprise above the average delay. Fig. 4 Bottom shows a correlation between the number of operations of airports and the corresponding thresholds (linear fit with slope $(7.37 \pm 5.11) \cdot 10^{-3}$, R-Square of 0.011). This indicates that the busiest the airport, the higher is the threshold, and the higher the additional delay needed to trigger a phase change. In other words, the big European hubs are more resilient to abnormal perturbations than small airports, a fact that may be explained in terms of available resources.

Once a set of thresholds is available for each airport, this information can be used to calculate how extreme events appear. Fig. 5 Left depicts the evolution of the number of extreme events through time, summed over all pairs of airports – note that, since the threshold is defined for airport pairs, an airport *A* may be in the disrupted phase with respect to airport *B*, and at the same time be in the normal phase with respect to another airport *C*. Extreme events seem to appear in burst, with a higher frequency during the summer – probably due to higher traffic levels. Fig. 5 Center and Right present box plots of the probability distributions of extreme events, respectively as a function of the day of the week and of the time of the day. In both cases, there are no significant changes in the average, as should be expected from the way delays time series have been made stationary. Nevertheless, there is a significant increase in the variability of the system during the peak hours, suggesting that the traffic volume has an important role in triggering the phase transition.

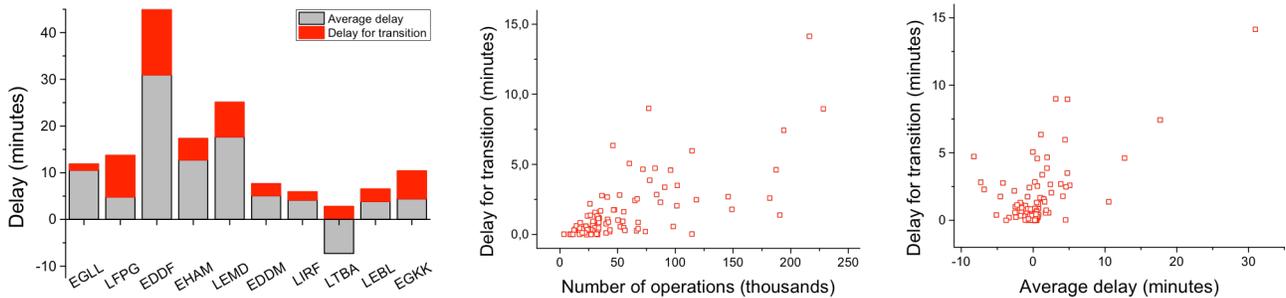

Figure 6. Additional delay required to trigger a phase change, for the 10 biggest airports (Left), and as a function of the number of operations (Center) and of the average delay (Right).

The thresholds of Fig. 4 can also be used to obtain an approximation of the average critical accumulated delay (within one hour) necessary to enter the phase where delays are out of control. If one denotes by $\tau_i$ the average threshold at airport *i*, the process used in Section III to make the time series stationary can be reversed as:

$$d_i^c \approx mean_j\left(\sigma^2(d_{ij})\right) * \tau_i + mean_j(d_{ij}),$$

where $d_i^c$ stands for the critical accumulated delay within one hour in the airport *i* leading to a phase transition. It is important to note that this is a first approximation: it merges average delays at different hours of the day (as represented by *j*), and the threshold $\tau_i$ itself is an average over all airports connected with *i*. In spite of this, $d_i^c$ is an interesting indicator of the stability of airports against severe disturbances.

The magnitude of the critical delay is represented in Fig. 6 Left for the 10 busiest airports, as the sum of the observed average delay (grey bars) and the additional delay to reach the threshold (red bars). If big airports were the most resilient of the system (see Fig. 4), Fig. 6 Left suggests that they are still very sensitive to disturbances; in some cases (*e.g.* EGLL) an increase as small as a 10% in the average delay is enough to switch to the disrupted phase. Fig. 6 Center and Right further depict the additional delay to reach the transition as a function of the number of operations[3] and of the average delay[4], respectively. In both cases, there is a small positive correlation, indicating that resilient airports are those with a high number of operations, and those that are accustomed to handling important delays.

V. DISCUSSION

In this contribution, we aimed at studying the topology created by delays propagation in Europe, through the application of two causality metrics. The first one is the celebrated Granger Causality, which detects if one time series can be explained in terms of a second one. The second metric is a newly proposed causality of extreme events, which specifically focuses on those parts of the time series that present abnormal values, and on the statistics of how such extreme events propagate.

The use of causality metrics allows us to describe the phenomenon of delay propagation from a global perspective, beyond the specific mechanisms driving it. If delays are measured at airports, this does not imply that airports are the only drivers. On the contrary, by measuring delays at the end of flights, all the elements involved in their generation, propagation and absorption are taken into account in an aggregated way: from ATFM, to late aircraft arrival and passengers connections. All of this, without requiring specific data on each one of these aspects.

---

[3] Linear fit with slope $5.1 \pm 1.2$, R-Square of 0.15.
[4] Linear fit with slope $0.35 \pm 0.11$, R-Square of 0.09.

The use of two distinct causality metrics allowed us to further differentiate between two different phases of the delays propagation. First, the standard one, which involves the propagation of small (expected) delays. Second, the propagation of extreme delays, *i.e.* those that are well beyond the normal expectations, and that are the result of unpredictable and severe disturbances (adverse weather, equipment failure, *etc.*). Intuitively, it would be expected that extreme delays may propagate in a different way than normal ones, as they are by definition outliers and the system is not prepared to handle them – *e.g.* airline buffers may be inadequate, the closure of a runway may unavoidably reduce the throughput, and so forth. Nevertheless, to the best of our knowledge, this is the first contribution tackling such problem in a quantitative way.

We relied on techniques from complex network theory to compare the structures created by both propagation processes. Extreme events are propagated through a denser network, *i.e.* airports influence more peers than under standard conditions; this is to be expected, as buffers (*e.g.* turn-around times between consecutive flights) are usually not designed to handle extreme delays. On the other hand, the extreme events propagation is less efficient, as it is more centered on small airports, and seldom threatens the whole system. Additionally, the combined use of both causality metrics allowed us to estimate the additional delay required for changing the phase of the propagation at each airport, *i.e.* to quantify when delays get out of control. Two types of airports have been identified as the weak links of the chain: small ones, probably due to their limited available resources to cope with unexpected situations; and those that are not accustomed to manage large delays. A few notable exceptions have also been highlighted: for instance, London Heathrow, which is extremely sensitive to unexpected delays, probably the result of continuously operating near the maximum capacity.

While this work is just an initial step, the results it presents will pave the way for a new understanding of the delay propagation phenomenon. They suggest that normal and extreme delays follow different dynamics, and that thus different mechanisms should be put in place for their mitigation. This topic should be the focus of future studies, also assessing the economical cost associated with both phases: in other words, it is more efficient to provide resources to absorb normal delays, abnormal ones, or to increase the average threshold of the system? Additionally, the quantification of the additional delays needed to change the propagation phase may be used as a resilience metric. In combination with standard metrics, this would allow to fully characterize each airport, both in terms of delays generated / absorbed under normal conditions, and of the capacity to retain a normal behavior under abnormal conditions.